\documentstyle[12pt,aaspp4]{article}

\newcommand{\fl}{\,{\rm erg\,s^{-1}cm^{-2}}}
\newcommand{\lum}{\,{\rm erg\,s^{-1}}}

\begin{document}

\null\vspace*{-3cm} \raisebox{3cm}{\hskip 9cm 
{\sf To appear in} {\sl The Astronomical Journal} }
\normalsize\vspace*{-2cm}\rm

\title{An X-ray Selected Galaxy Cluster at $z=1.26$
\altaffilmark{1,}\altaffilmark{2}}

\altaffiltext{1}{Based in part on observations obtained at the W.M.\ Keck
Observatory}
\altaffiltext{2}{Based in part on observations obtained at the Kitt Peak National Observatory}

\author{Piero\ Rosati\altaffilmark{4}} 
\affil{ESO -- European Southern Observatory, \\D-85748 Garching
  bei M\"unchen (Germany)} 
\authoremail{prosati@eso.org}

\author{S.A.\ Stanford\altaffilmark{3}} 
\affil{Institute of
Geophysics and Planetary Physics, Lawrence Livermore National
Laboratory, Livermore, CA 94550} 
\authoremail{adam@igpp.llnl.gov}
\altaffiltext{3}{Physics Department, University of California at Davis, 
Davis, CA  95616.}
 
\author{Peter R.\ Eisenhardt\altaffilmark{4}} \affil{Jet Propulsion
Laboratory, California Institute of Technology, Pasadena, CA 91109}
\authoremail{prme@kromos.jpl.nasa.gov}

\author{Richard Elston\altaffilmark{4}} \affil{Department of
Astronomy, University of Florida, Gainesville, FL 32611 }
\authoremail{elston@astro.ufl.edu}

\altaffiltext{4}{Visiting Astronomer, Kitt Peak National
Observatory, National Optical Astronomy Observatories, which is
operated by the Association of Universities for Research in Astronomy,
Inc., under cooperative agreement with the National Science
Foundation.}

\author{Hyron Spinrad, Daniel Stern}
\affil{Astronomy Department, University of California, 
Berkeley, CA  94720}
\authoremail{spinrad@bigz.berkeley.edu, dan@bigz.berkeley.edu}

\and
\author{Arjun Dey}
\affil{National Optical Astronomy Observatories, Tucson, AZ 85726-6732}

\authoremail{dey@noao.edu}

\begin{abstract}
We report the discovery of an X-ray luminous galaxy cluster at $z =
1.26$.  RXJ0848.9+4452 was selected as an X-ray cluster
candidate in the ROSAT Deep Cluster Survey, on the basis of its
spatial extent.  Deep optical and near-IR imaging have revealed a
galaxy overdensity around the peak of the X-ray emission, with a
significant excess of red objects with $J-K$ colors typical of
elliptical galaxies at $z>1$.  Spectroscopic observations at the Keck
II telescope have secured  6 galaxy redshifts in the range 
$1.257<z<1.267$ ($<z>=1.261$), within a 35 arcsec radius around the
peak X--ray emission.  This system lies only 4.2 arcmin away ($5.0\,
{\rm h}^{-1}_{50}$ comoving Mpc, $q_0=0.5$) from the galaxy cluster
ClG~J0848+4453, which was identified by Stanford et al.\ (1997) at
$z=1.273$ in a near-IR field galaxy survey, and is also known to be
X-ray luminous.  Assuming that the X-ray emission is entirely due to
hot intra-cluster gas, both these systems have similar rest frame
luminosities $L_X \approx 1\times 10^{44}$ ergs s$^{-1}$ (0.5--2.0 keV
band). In combination with our spectrophotometric data for the entire
30 arcmin$^2$ field, 
this suggests the presence of a superstructure,
consisting of two collapsed, possibly virialized clusters, the first
detected to date at $z>1$.

\end{abstract}


\section{Introduction}

Over the last few years significant progress has been made in the
search for distant galaxy clusters as well as in the study of their
galaxy populations.  The existence of massive, virialized systems at
very high redshifts has a direct bearing on theories of structure
formation (e.g., Eke et al. 1996).  
Finding clusters at redshift $z>0.5$ has become routine in
serendipitous searches in the X-ray, based on deep ROSAT-PSPC 
observations.  Near-infrared surveys are now reaching the necessary
depth and area to push these searches to redshifts $z>1$ 
(e.g.,\ Stanford et al.\ 1997).  These studies have provided
new constraints on both the evolution of the cluster abundance out to
$z\simeq 0.8$ (see Rosati 1998 for a review) and the evolution of 
early-type galaxies over a substantial look-back time (e.g.,\ Stanford
1998).

The study of the spectrophotometric properties of cluster galaxies at
large redshifts provides a powerful means of discriminating between
scenarios for the formation of elliptical galaxies. Predictions of the
color-magnitude distribution of cluster early-types based on
hierarchical models of galaxy formation (Kauffmann 1996; Kauffmann \&
Charlot 1998)  differ at $z > 1$ from models in which ellipticals
formed at high-$z$ in a monolithic collapse (Eggen et al.\ 1962). This
is the result of different star formation histories predicted in the
two scenarios.  In the former, star formation activity is modulated by
interactions and merging up to relatively late times, whereas in the
single burst scenario the original stellar population evolves
passively for a large fraction of the Hubble time.  The evolution of
the IR-optical colors, the slope of the color-magnitude relation, and
the amount of its scatter are important diagnostics for constraining
the mode and epoch of formation of the E/S0 galaxies (Bower, Lucey, \&
Ellis 1992; Kodama \& Arimoto 1997; Ellis et al.\ 1997; Stanford,
Eisenhardt, \& Dickinson 1998).  All these studies, which to date have
mainly concentrated on clusters at $z\lesssim 1$, suggest that
early-type galaxies  form at $z_f >3$, with a high degree of
synchronism in their initial epoch of star formation activity.
Identifying
clusters at $z>1$ provides a valuable sample of galaxies at early
cosmic epoch.  The rest frame ultraviolet spectra can be used to
age date the stellar populations, providing another measure of the
formation epoch of early-type galaxies (e.g., Dunlop et al. 1996; Spinrad
et al. 1997).

Near-infrared imaging is essential at these large redshifts to
compensate the k-correction which significantly dims the dominant
population of early-type cluster galaxies at observed optical
wavelengths. Stanford et al.\ 1997 (hereafter S97) have shown that
optical-infrared colors can be used to considerably enhance the
contrast of the cluster galaxies against the field galaxy population
and, at the same time, to estimate the cluster
redshift.  By using this method S97 detected a red clump of galaxies
with a narrow $J-K$ color distribution in a field galaxy survey
covering $\sim\! 100$ arcmin$^2$ (Elston et al.\ 1999). Follow-up
spectroscopy with the Keck telescope secured 10 galaxy redshifts at
$<z>=1.273$ with a velocity dispersion of $640\pm 90$ km/s.  An
a posteriori analysis of a deep ROSAT-PSPC archival pointing revealed a
corresponding low-surface brightness X-ray emission at the cluster
position, most likely the result of hot intra-cluster gas with an X-ray
luminosity of $L_X[0.5-2.0\ {\rm keV}]=(0.8\pm 0.3)\times 
10^{44}\lum$, i.e.\ typical
of a moderately rich cluster. This system, ClG~J0848+4453, is
currently the most distant X-ray emitting cluster found
serendipitously in a field survey. 
Targeted searches for clusters around high-redshift AGN 
have led to several positive identifications (e.g., Crawford \& 
Fabian 1996; Pascarelle et al. 1996; Dickinson 1997; Hall \& 
Green 1998).
 Follow-up deep ROSAT
imaging has also revealed the existence of diffuse X-ray emission
around several 3CR sources out to $z\simeq 1.8$ (Dickinson et
al.\ 1999). The rarity of powerful radio galaxies, however, make these
studies unsuitable for assessing the cluster abundance, and perhaps
unrepresentative of normal cluster environments.

X-ray imaging is also a well established method to search for
bona-fide distant clusters (e.g., Gioia et al.\ 1990, Gioia \& Luppino, 1994).
With the advent 
of the ROSAT-PSPC, with its unprecedented sensitivity and spatial
resolution, archival searches for extended X-ray sources have become a
very efficient method to construct large, homogeneous samples of
galaxy clusters out to at least $z\simeq 0.8$ (e.g.,\ Rosati et
al.\ 1995, 1998; Scharf et al.\ 1997; Collins et al.\ 1997; Vihklinin et
al.\ 1998).  The ROSAT Deep Cluster Survey (RDCS) (Rosati et al.\ 1998
(R98)), has shown no evidence of a decline in the space density of 
galaxy clusters of X-ray luminosity $L_X\lesssim L_X^\ast$ over a wide
redshift range, $0.2<z<0.8$.  The fact that the bulk of the X-ray
cluster population is not evolving significantly out to this large
redshift increases the chances of finding clusters beyond redshift
one, since $L_X^\ast$ clusters ($\approx 4\times 10^{44}\lum$
in the [0.5-2.0] keV band, roughly the Coma cluster) can be detected at $z>1$ 
as extended X-ray sources in deep ROSAT pointed observations, provided
that the X-ray surface brightness profile does not evolve
significantly (see also Hasinger et al. 1999).

To improve the success rate of identifying very high redshift clusters
from deep X-ray surveys, we have begun a program of near-IR imaging of
unidentified faint candidates in the RDCS.  In this paper, we describe
the imaging and spectroscopic follow-up observations of the first of
these X-ray faint candidates, RXJ0848.9+4452, which has led to the
discovery of a cluster at $z=1.261$. The corresponding X-ray source is
located in the same PSPC field (the ``Lynx field'') as the cluster discovered
 by S97, and lies only 4.2 arcmin away from ClG~J0848+4453 at
$z=1.273$.  By combining these new data with the spectrophotometric
data of S97, we find strong evidence for the existence of a superstructure 
at $z\approx 1.265$ consisting of two collapsed, possibly
virialized clusters.
Unless otherwise stated, we adopt the parameters
$H_0=50$ km s$^{-1}$ Mpc$^{-1}$, $q_0=0.5$.
 
\section{The X-ray Selection and Analysis}
The RDCS was designed to compile a large, X-ray flux-limited sample of
galaxy clusters, selected via a serendipitous search for extended
X-ray sources in ROSAT-PSPC deep pointed observations (Rosati et
al.\ 1995, 1998).  The depth and the solid angle of the survey were
chosen to probe an adequate range of X-ray luminosities over a large
redshift baseline.  Approximately 160 candidates were selected down to
the flux limit of 
$f_{-14} = 1$ (where $f_{-14} = F_X(0.5-2.0\ {\rm keV})/10^{-14}\fl$), 
over a total
area of 50 deg$^2$, using a wavelet-based detection technique. This
technique is particularly efficient in discriminating between
point-like and extended, low surface brightness sources
(Rosati et al.\ 1995).

Optical follow-up observations for the cluster candidates consisted
primarily of optical imaging in the $I$-band, followed by multi-object
spectroscopy with 4m class telescopes at NOAO and ESO. To date, more
than 100 new clusters or groups have been spectroscopically confirmed, and
about a quarter of these lie at $z>0.5$. The spectroscopic
identification is 85\% complete to $f_{-14}=3$. To this
limiting flux the completeness level and the selection function are
well understood. At fainter fluxes the effective sky coverage
pregressively decreases. In addition, the X-ray identification becomes
more difficult due to the increasing confusion and the low
signal-to-noise ratio of the X-ray sources which makes it more
difficult to discriminate between point-like and extended
sources. Below $f_{-14}=2$ the completeness level can be as low as
50\%. However, it is in these lowest flux bins that the most distant
clusters of the survey are expected to lie. If one assumes that the
evolutionary trend in the cluster population continues past $z=1$, the
observed X-ray luminosity function, $N(L_X,z)$, of R98 can be
extrapolated to predict that up to a dozen clusters at $z\gtrsim 1$
remain to be identified in the RDCS, which covers about 15 deg$^2$ at
$f_{-14}=3$. This prediction conservatively takes into account the
evolution of the bright end ($L_X>L_X^\ast$), in keeping with the
results of Henry et al.\ (1992), and the survey incompleteness near the
flux limit.

To date, moderately deep $I$-band imaging of several candidates with
$f_{-14}<3$ has shown only marginally significant galaxy overdensities
in the best cases. Some of these candidates could be X-ray sources of
a different nature, rather than galaxy clusters. To increase the
efficiency of cluster identification, we have initiated a program of
$J$ and $K$-band imaging of these faint unidentified candidates in the
deepest archival fields of the RDCS.  The Lynx field, centered at
$\alpha = 08^h 49^m 12\,\fs0, \delta = +44\arcdeg 50\arcmin 24\arcsec$
(J2000), was observed for 64.3 ksec with the PSPC in two pointings
(rp90009A00,A01). The X-ray data were processed as described in Rosati
et al.\ 1995. The object detection and classification yielded 3
cluster candidates in the central area of the detector ($\theta
<15\arcmin$) (see Table 1). The brightest one was identified as a
cluster at $z=0.571$, using the CryoCam spectrograph at the KPNO 4m
telescope.  A 1200 second $I$ band image of the two faint remaining
candidates, obtained at the KPNO 4m prime focus, failed to show a
strong enhancement in the projected galaxy density at either of the
peaks of the X-ray emission.

A 28 arcmin$^2$ region in the Lynx field was also part of a deep
multicolor optical-IR survey by Elston et al.\ (1999). The faintest
source in Table 1, RXJ0848.6+4453, was identified by S97 as the X-ray
counterpart of ClG~J0848+4453 at $z=1.273$.  In the following, we will
focus on RXJ0848.9+4452, the cluster candidate which lies at the
extreme southeast corner of the Lynx field in the Elston et al.\
near-IR survey.

\begin{figure}
\epsscale{0.9}
\plotone{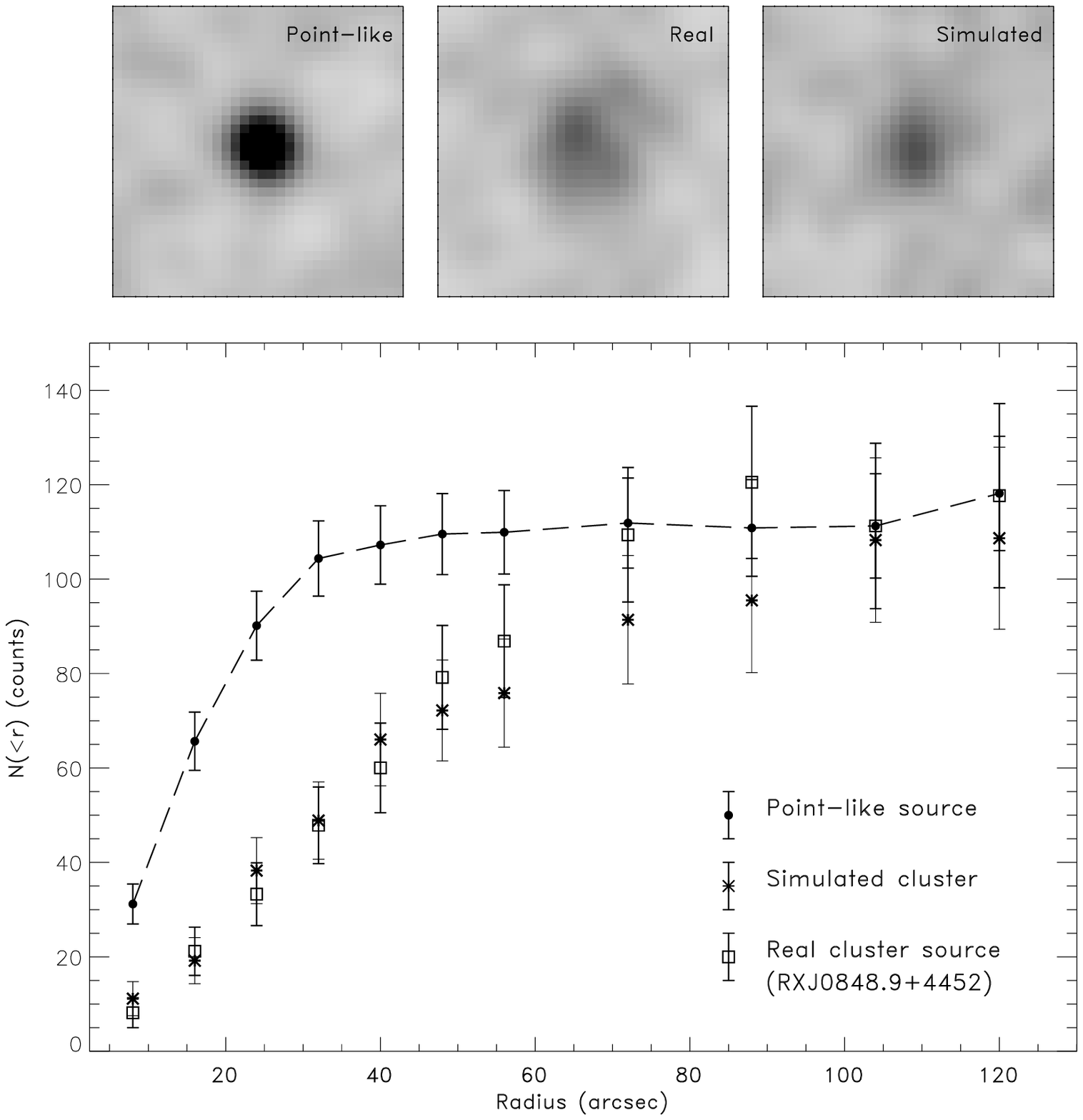}
\caption{Comparison between the bidimensional and the radial
cumulative surface brightness of a point-like source in the Lynx field, 
a simulated extended source and the X-ray source RXJ0848.9+4452.}  
\epsscale{1.0}
\label{f:sb}
\end{figure}

In Figure~\ref{f:sb}, we show a comparison between the bidimensional
and radial cumulative X-ray surface brightness of RXJ0848.9+4452 with a
nearby point-like source scaled down by a factor 1.8 to account for the
flux ratio of the two sources.  The latter is
located 2 arcmin north ($\theta <4.9\arcmin$) and has been identified
as a QSO at $z=0.575$ in spectroscopic observations carried out at the
KPNO 4m telescope.  We use this point-like source to register the PSPC
X-ray image onto the optical images.  This resulted in a 10\arcsec\
shift from the nominal position, which is not unusual for aspect
solution errors of the PSPC.  The statistical $1\sigma$ residual error
in the position of the X-ray centroid is about 5\arcsec\,.  The
conversion factor from counts to unabsorbed X-ray flux was calculated
using the galactic HI column density in the Lynx field, $N_H=3\times
10^{20} {\rm cm}^{-2}$, and assuming a thermal spectrum with $T=6$ keV
for the source. The source counts are integrated over an aperture of
$2\arcmin\ \approx 1 \, h_{50}^{-1}$ Mpc radius.  This aperture
encircles $\sim\! 86\%$ of the total flux for the profile parameters
given below.

The first and second top panels in Figure~\ref{f:sb} are cut-outs
extracted from the PSPC image in the [0.5-2.0]~keV band with 8\arcsec\
pixels. The third panel shows a simulated cluster which has been
obtained assuming a King profile with core radius of 200 kpc,
$\beta=0.7$, and the observed X-ray flux of $ 1.8 \times 10^{-14}
\fl$.  The X-ray surface brightness count distribution is then
redshifted to $z=1.26$ and overlaid onto the measured Poissonian
background in the Lynx field of 0.37 counts/pixel.  This simulation
shows that, within the errors due to the low signal-to-noise, typical
cluster parameters match the surface brightness distribution of
RXJ0848.9+4452. Furthermore, the growth curves show that the cluster
candidate source is clearly extended when compared with a nearby
point-like source, as also revealed by the statistical
characterization provided by the wavelet algorithm (Table 1).

Given the limited resolution of the PSPC and the faintness of the
X-ray source, we cannot rule out the possibility from the X-ray
properties alone that the X-ray emission is the result of source
confusion, or is partially contaminated by one or more AGN. The X-ray
spectral information on the X-ray source is very poor, although no
emission is detected below 0.5 keV, consistent with a thermal cluster
spectrum.  As we will discuss below, however, the optical-IR imaging
and spectroscopic follow-up observations suggest that the diffuse
X-ray emission is most likely due to hot intracluster gas trapped in a
gravitationally bound structure.

\section{Imaging and Spectroscopy}

\begin{figure}
\epsscale{0.8}
\plotone{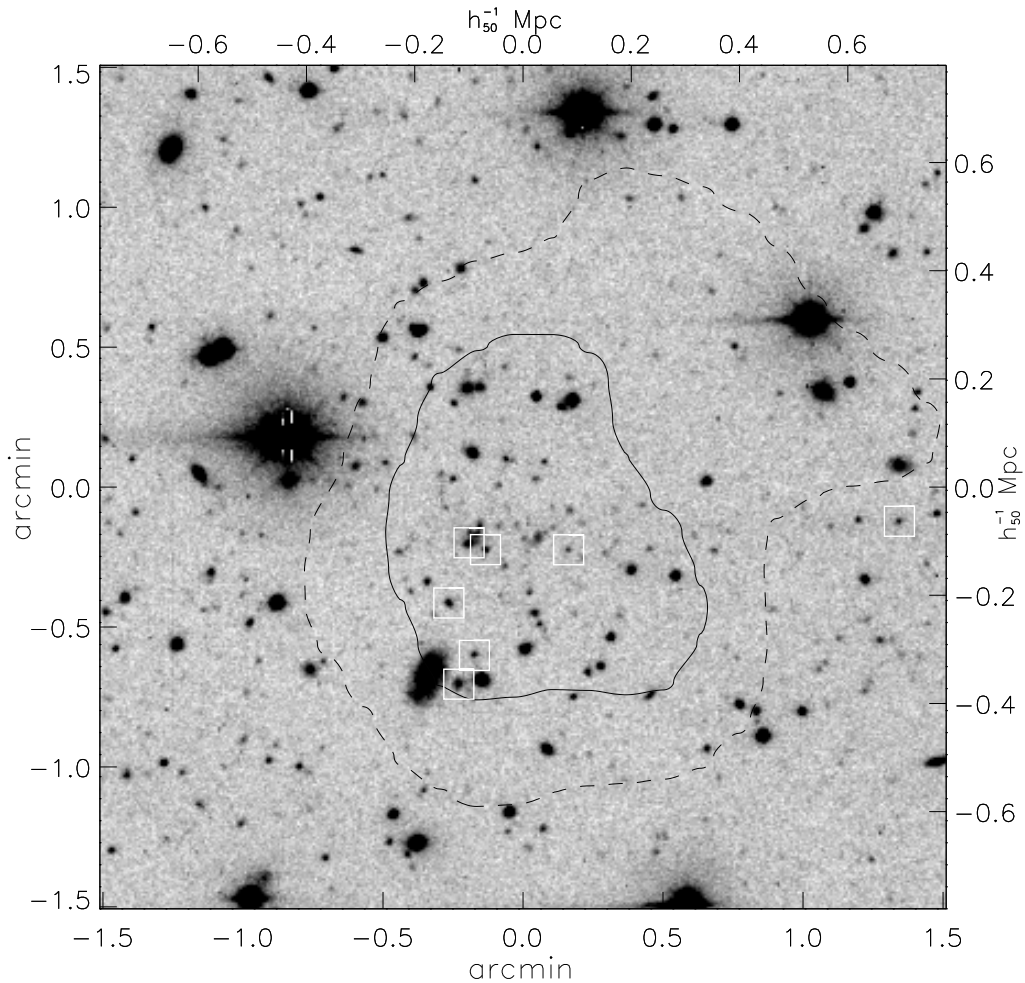}
\caption{$I$-band image of RXJ0848.9+4452 with the PSPC X-ray contours
at 3 and 7 $\sigma$ overlaid, centered at coordinates
$\alpha = 08^h 48^m 42\,\fs0, \delta = +44\arcdeg 54\arcmin 00\arcsec$ (J2000).
The image was obtained at the KPNO-PF 4m
telescope with 9600 sec of integration. Spectroscopically-confirmed
members at $1.257<z<1.268$ are marked by boxes. The right and top axes give
the physical linear scale at $z=1.26$.}
\label{f:Ibandima}
\end{figure}

Deep optical imaging of the Lynx field in $BRIz$ bands was carried out
by Elston et al.\ (1999) at the 4m telescope of the Kitt Peak National
Observatory using the PFCCD/T2KB which covers $16\times 16$ arcmin with
0.48\arcsec\ pixels.  A deep $I$-band image, taken as part of this
survey, is shown in Figure~\ref{f:Ibandima} with the X-ray contours
overlaid. Although an overdensity of galaxies with $I=21-23$ is visible
around the X-ray peak ($1\sigma$ significant above the background
counts), optical data alone cannot distinguish between possible cluster
galaxies and the field galaxy population.

\subsection{Followup IR imaging}

\begin{figure}
\epsscale{0.8}
\plotone{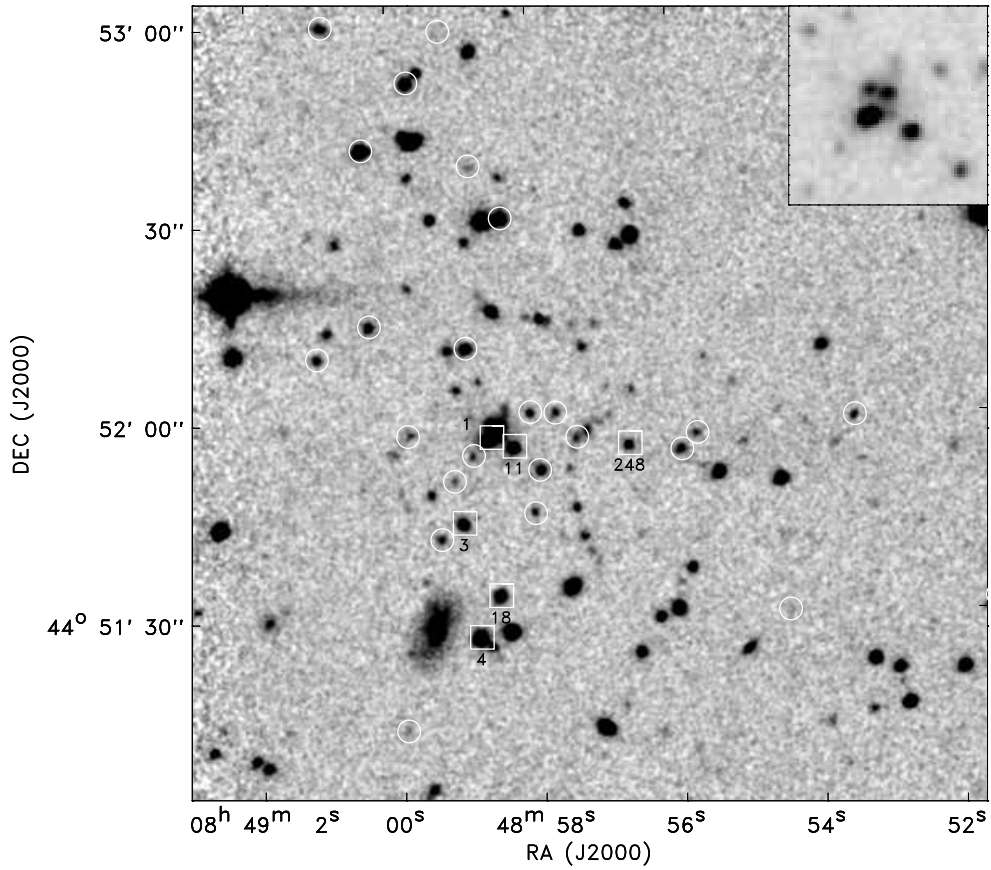}
\caption{$K$-band image of the central area of
fig.~\protect\ref{f:Ibandima}.
 Spectroscopically-confirmed members at $1.257<z<1.268$ are marked by
boxes along with their IDs; possible cluster members with
$1.8<J-K<2.1$ are circled. The inset shows a blow-up of the cluster
core covering $\sim\! 150\,{\rm h}^{-1}_{50}$ kpc at $z=1.26$.}
\label{f:Kbandima}
\end{figure}

Since the area around RXJ0848.9+4452 was only partially covered by the
near-IR survey of Elston et al., we obtained $J$ and $K_s$ imaging at
the Palomar 200" telescope with the Prime-Focus Infrared Camera
(Jarrett et al.\ 1994).  This camera provides a 2.1\arcmin\ field of
view with 0.494\arcsec\ pixels.  The source was observed on 23 March
1998 through cirrus and in photometric conditions on 24 March 1998.
The flux scale was calibrated using observations of three UKIRT
standard stars at similar airmass which bracketed the RXJ0848.9+4452
observations on 24 March 1998.  The data were taken using a sequence
of dither motions in both axes with a typical amplitude of 15\arcsec,
and a dwell time between dithers of 40 seconds at $J$ and from 15 to
30 seconds at $K_s$.  The data were linearized using an empirically
measured linearity curve taken during the observing run, and reduced
using DIMSUM\footnote{ Deep Infrared Mosaicing Software, a package of
IRAF scripts available at ftp://iraf.noao.edu/contrib/dimsumV2}.  The total
integration times and FWHM's of the resulting images are 3360~s and
1.1" at $J$, and 5600~s and 0.9" at $K_s$.
The $K_s$ image of the RXJ0848.9+4452 field is shown in 
Figure~\ref{f:Kbandima}.

A catalog of objects in the $K$-band image was obtained using FOCAS
(Valdes et al.\ 1982), as revised by Adelberger (1996).  The $K$-band
image was smoothed by a small amount to match the 1.1 arcsec
seeing in the $J$-band image.  Objects were detected with the
requirement that contiguous pixels covering an area of 0.94 arcsec$^2$
must be 3 $\sigma$ above the background.  All detected objects were
inspected visually to eliminate false detections.  The catalog is 90\%
complete to $K = 21.5$ in a 2 arcsec aperture.  The catalog was then
applied to the $RIJ$ band images, which had been geometrically
transformed to the $K$-band image, to obtain matched aperture
photometry in those bands.

\begin{figure}
\epsscale{0.9}
\plotone{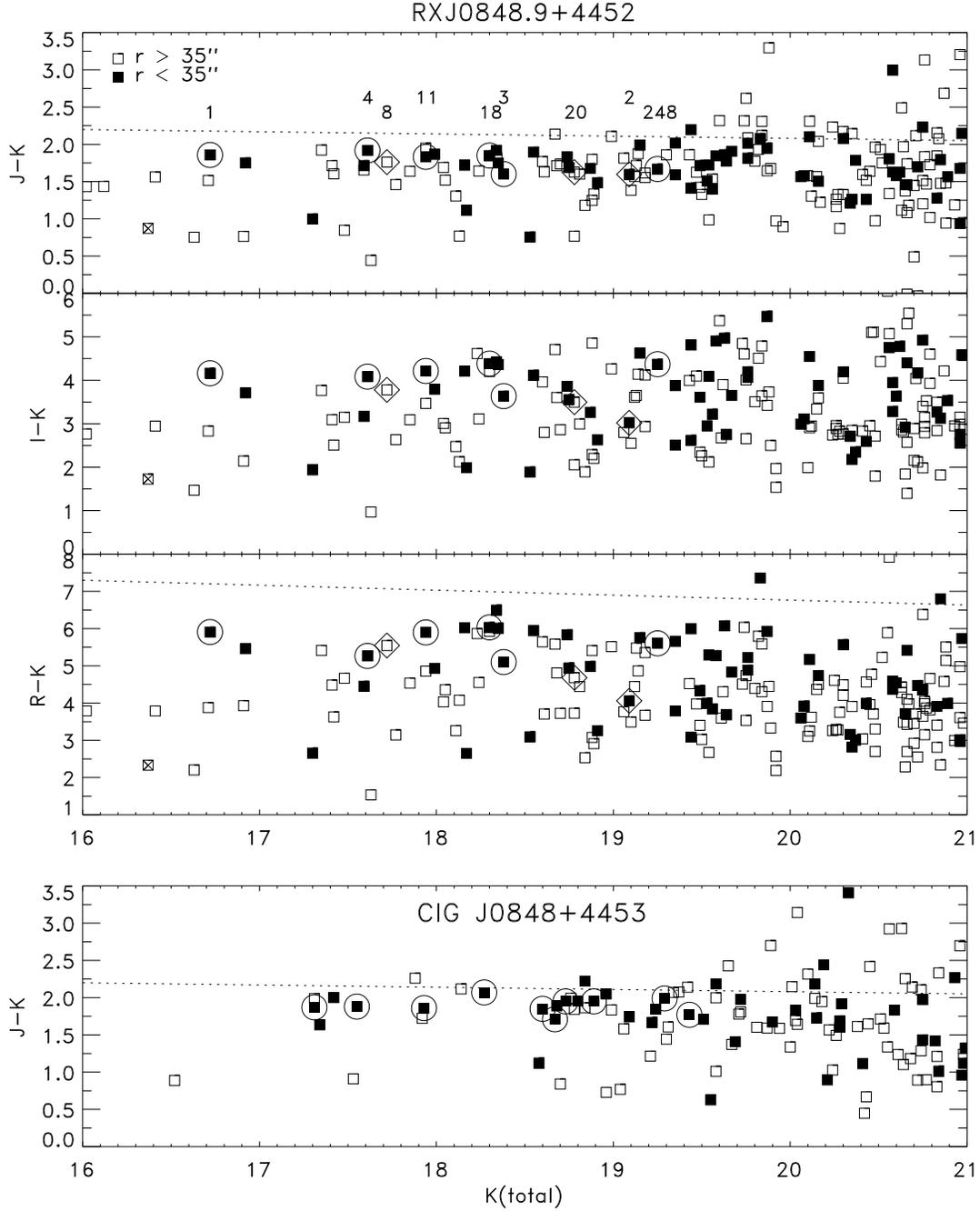} \medskip
\caption{{\sl Top}: Color magnitude diagram of the 2\arcmin\ field around RXJ0848.9+4452.
Open (filled) squares indicate objects lying outside (inside) a
circular area of radius $35\arcsec\ = 300 {\rm h}_{50}^{-1}$ centered
at the peak of the X-ray source.  The circled objects are the
spectroscopic members. Boxed objects are foreground galaxies with
spectra.  The dashed lines represent a predicted no-evolution locus
for early-type galaxies based on the Coma cluster. 
{\sl Bottom}: Color magnitude diagram of the 9 arcmin$^2$ area around
ClG~J0848+4453. The bottom panel updates Fig.\ 4 of Stanford et al.\ 1997
including ten spectroscopic members which have been confirmed to date. 
}
\label{f:cmdiagram}
\end{figure}

Figure~\ref{f:cmdiagram} shows the resulting color-magnitude diagrams
for all objects in a 2 arcmin$^2$ area around RXJ0848.9+4452.  The
top panel shows that the $J-K$ color distribution of objects falling
within a circular area of 35\arcsec\ radius, centered on the X-ray
source, is significantly skewed toward the red, with a maximum at
$J-K\approx 1.85$. This spatial segregation of red objects around the
X-ray centroid is more evident in Figure~\ref{f:Kbandima}, where the
objects belonging to the red sequence, $1.8<J-K<2.1$, are marked
on the $K$ band image of the central area.  This overdensity of red
objects is comparable with that found by S97 around ClG~J0848+4453.
There are 21 objects with $1.8<J-K<2.1$ down to $K=20$ which fall within
a circular area of radius 60\arcsec\ around RXJ0848.9+4452, i.e a
density of 6.7 arcmin$^{-2}$, in contrast to the average density of
3.3 arcmin$^{-2}$ over the
entire $\approx 100$ arcmin$^2$ area of the Elston et al. survey
(Eisenhardt et al. 1998). 
The $J-K$ vs $K$ diagram for ClG~J0848+4453 is
shown for comparison in the bottom panel of Fig.~\ref{f:cmdiagram}.  
A zeropoint error that
results in the $J-K$ colors of ClG~J0848+4453 becoming bluer
by 0.1 mag was recently discovered in the JK photometry reported in S97.
This error has been corrected in the lower panel of Fig.~\ref{f:cmdiagram}.

\subsection{Keck Spectroscopy}

Spectroscopic observations of selected galaxies in a 2\arcmin\ region
around RXJ0848.9+4452 were obtained using the Low Resolution Imaging
Spectrometer (LRIS; Oke et al.\ 1995) on the Keck II telescope.
Objects were assigned slits based on their $J-K$ colors and optical-IR
magnitudes.  Spectra were obtained using the 400 l mm$^{-1}$ grating
which is blazed at 8500\AA, covering the $\sim 6000-9800$ \AA~region.
The dispersion of $\sim$1.8 \AA~pixel$^{-1}$ resulted in a spectral
resolution of 9 \AA~ as measured by the FWHM of emission lines in arc
lamp spectra.  Usually each mask was observed for 5-6 1800~s
exposures, with small spatial offsets along the long axis of the
slitlets.  Three slitmasks in the Lynx field were used to obtain
spectra on 20-21 January 1998 UT, 18 February 1998 UT, and 28-29 March
1998 UT.

The slitmask data were separated into individual slitlet spectra and
then reduced using standard longslit techniques.  The exposures for
each slitlet were reduced separately and then coadded.  
One--dimensional spectra were extracted for each of the targeted
objects.  Wavelength calibration of the 1-D spectra was obtained from
arc lamp exposures taken immediately after the object exposures.  A
relative flux calibration was obtained from a longslit observation of
the standard stars HZ 44 and G191B2B (Massey et al. 1988, 1990)
with the 400 l mm$^{-1}$ grating.
While these spectra do not straightforwardly yield an absolute flux
calibration of the slit mask data, the relative calibration of the
spectral shapes is accurate.

\begin{figure}
\epsscale{0.6}
\plotone{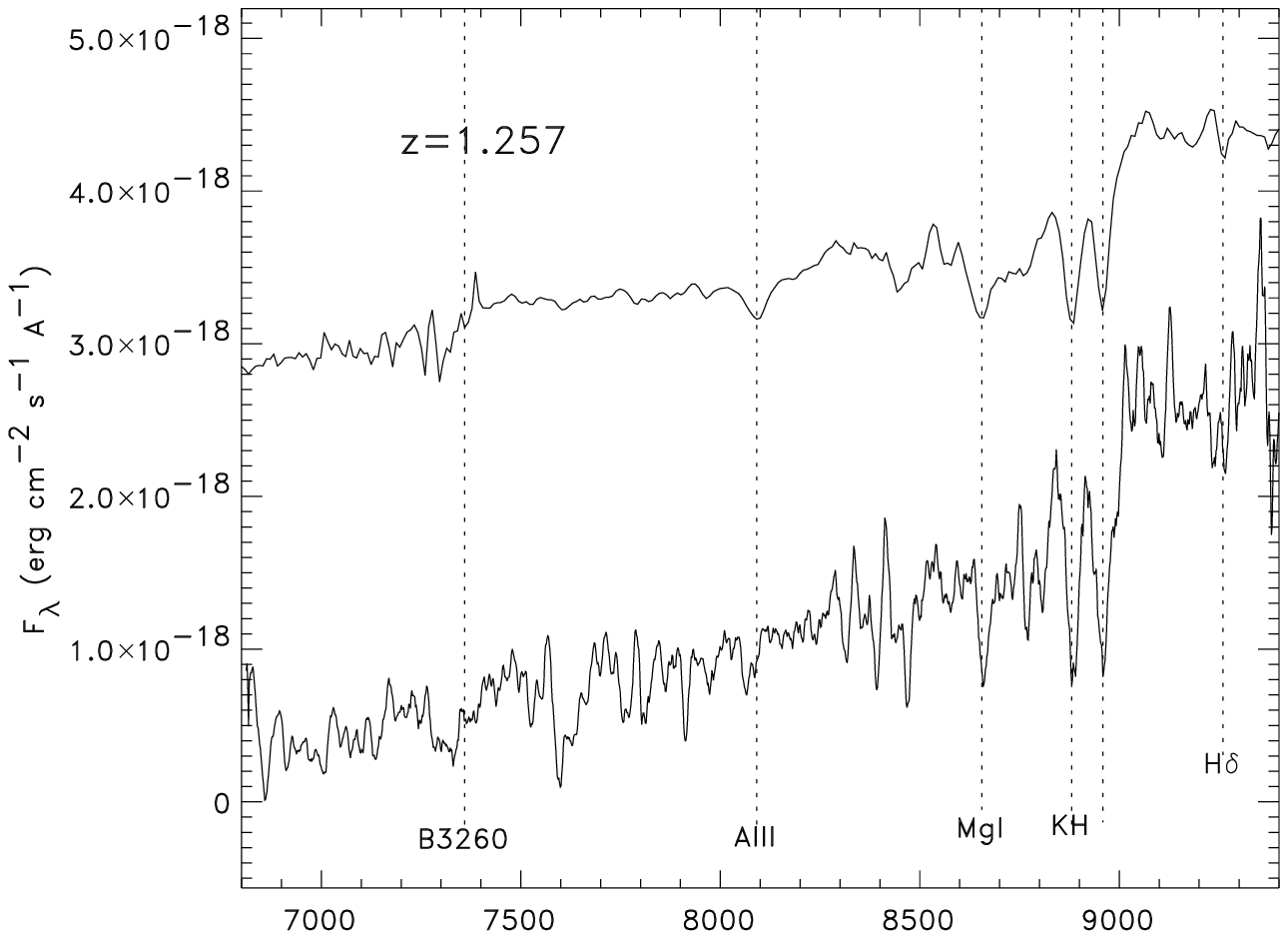} \\
\plotone{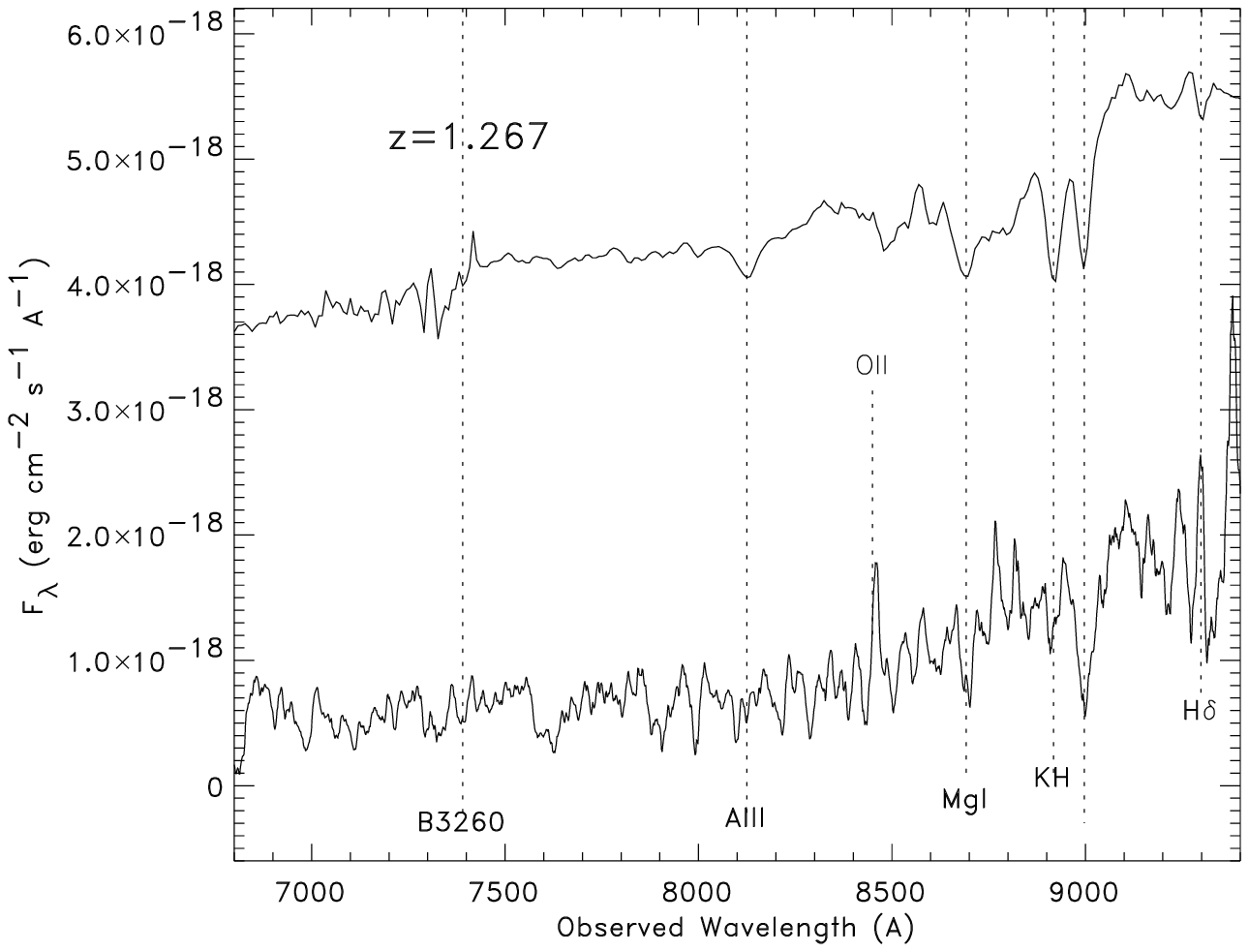}
\caption{Optical spectrum of Objects 1 and 4 in RXJ0848.9+4452 smoothed by a
9 pixel boxcar (3 hours integration). A template of a local elliptical
galaxy used to measure the redshift is also shown.}
\label{f:spectra}
\end{figure}

Redshifts were calculated by cross-correlating the spectra with an E
template from Kinney et al.\ (1996) using the IRAF Package RVSAO/XCSAO
(Kurtz et al.\ 1991). The redshift measurement was based on matching
major absorption features, such as Ca II H+K, Mg I $\lambda$~3830 and
$\lambda$~2852, and Mg II $\lambda~2800$. The cross correlation is
also sensitive to spectral breaks such as D4000 and B2900.  Only one
object (ID~\#4) shows weak [OII] emission, and the remaining galaxies
have spectra similar to the local E template. 
Two examples of spectra (ID~\#1,\#4), rebinned by a factor of 9, are shown in Figure~\ref{f:spectra}.

Five galaxies were found to have redshifts in the range
$1.257<z<1.267$ (an additional galaxy, ID~\#248, has a less certain
redshift),  i.e. a relative velocity
$\Delta v\sim\! 1300$ km/s.  This range covers twice the velocity
dispersion estimated by S97 for ClG~J0848+4453.

\section{Discussion}

The spectrophotometric properties of all the targets around
RXJ0848.6+4453 are summarized in Table 2. Two additional galaxies (not
listed in the Table, one visible in fig.~\ref{f:Ibandima}) found by
S97 at $z=1.268$ and $z=1.265$ lie within 2\arcmin\ of the central
condensation (marked by object \#1).  Our
spectrophotometric data, combined with the evidence of a significant
enhancement in the density of red galaxies against the field,
coinciding with the diffuse X-ray emission, strongly suggest the
presence of a gravitationally bound structure at $z=1.261\pm 0.005$.
Assuming that the X-ray emission is not contaminated by faint AGN and
is entirely due to hot intra-cluster gas trapped in the potential well
of this structure, we obtain an X-ray luminosity of $1.5\times
10^{44}\lum$ in the rest frame [0.5--2.0] keV band (the k-correction
amounts to 1.25 for a thermal spectrum at $T=6$ keV), typical of a
moderately rich cluster.

\subsection{A Superstructure at $z\simeq 1.27$} 
\begin{figure}
\epsscale{0.9}
\plotone{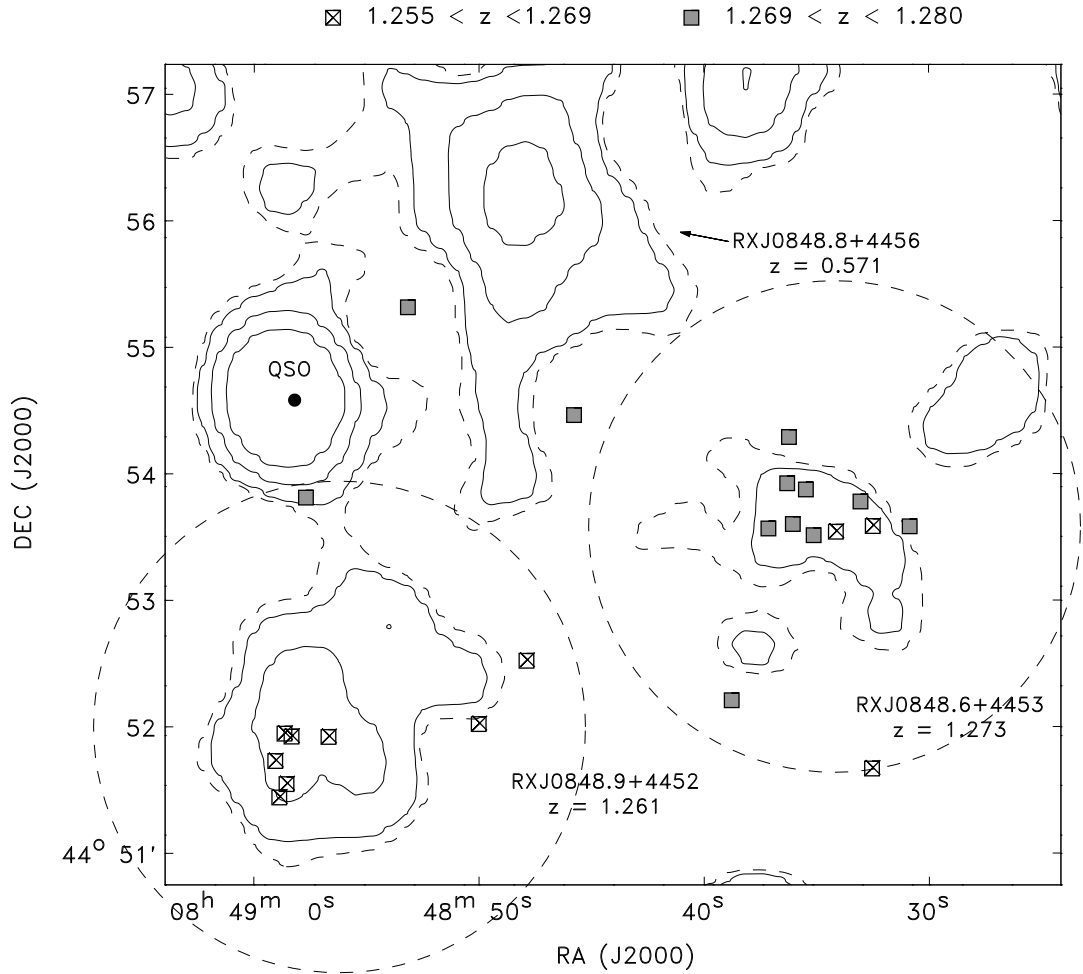}
\caption{Lynx field with overlaid ROSAT-PSPC contours and spectroscopic
 members. 
The circles are centered at the
 X-ray centroids of the two clusters with a radius of $116\arcsec\ =
 1\, {\rm h}^{-1}_{50}$ Mpc at $z=1.27$. Also indicated are the QSO and 
 the foreground  cluster both at $z=0.57$ (top). }
\label{f:xcont_obj}
\end{figure}

\begin{figure}
\epsscale{0.8}
\plotone{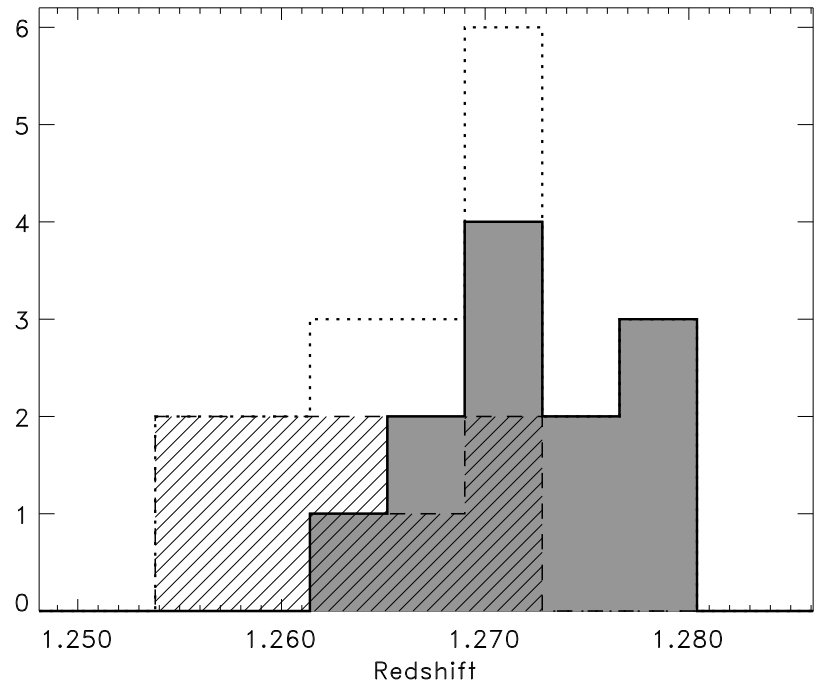}
\caption{Histograms of the redshifts of cluster members drawn from the
circular areas in Figure~\protect\ref{f:xcont_obj}, where the hatched
area represents RXJ0848.9+4452 and the solid area represents
ClG~J0848+4453.  The dotted line shows the sum of all members in the
two clusters with $1.25<z< 1.28$.
}
\label{f:hist_z}
\end{figure}

We show in Figure~\ref{f:xcont_obj} a map of the identified X-ray
sources in a 40 arcmin$^2$ area of the Lynx field, as well as the
galaxies with spectroscopic redshifts in the range $1.25<z<1.28$,
which include the spectroscopic data presented by S97.  A composite
$BIK$ image of the same area is shown in Fig.~\ref{f:bik}.  The
redshift histogram of all the spectroscopic members in the two
clusters is shown in Figure~\ref{f:hist_z}.  RXJ0848.9+4452 lies only
4.2\arcmin\ away from ClG~J0848+4453 (alias RXJ0848.6+4453), i.e.\ at
a projected physical distance of $2.2\, {\rm h}^{-1}_{50}$ Mpc at that
redshift ($5.0\, {\rm h}^{-1}_{50}$ comoving Mpc).  These data provide
evidence for the presence of a superstructure at $z\simeq 1.27$
consisting of two collapsed systems, which are likely in an advanced
dynamical state, considering that X-ray emitting gas had the time to
thermalize in their potential wells.

Inspection of Figure~\ref{f:xcont_obj} also shows some evidence of a
kinematical segregation of the spectroscopic members in the two
systems, which requires confirmation with additional spectroscopy.
 A relative radial velocity of $\sim\! 1000$ km/s has
been found in similar X-ray clusters pairs found at $z=0.55$ (Hughes et
al.\ 1995, Connolly et al.\ 1996) and in the RDCS at lower redshifts.

The X-ray properties of the two systems are similar, although the new
cluster has an X-ray luminosity about twice that of ClG~J0848+4453
(Table 1).  If the estimate of the X-ray fluxes are not affected by
source confusion, the higher X-ray luminosity of RXJ0848.9+4452,
together with its near-IR appearance showing a more compact
distribution of red galaxies compared with its $z=1.27$ companion,
suggests that it may be in a more advanced dynamical state. 
Detailed analysis of the morphology of the X-ray emission is not
possible with the current ROSAT-PSPC data and signal-to-noise. A 
deep ACIS exposure scheduled with AXAF in A01 will be able to
unambiguously establish the presence of X-ray emitting gas,
characterize the thermal state of these two clusters, and permit a
measurement of the gas temperatures and metal abundances.
Without these additional data it is difficult to obtain a reliable
estimate of the cluster mass. Using a larger number of spectroscopic
members, S97 estimated the velocity dispersion of  ClG~J0848+4453 to be
$640$ km/s, which translates into a mass $M(r<3 {\rm
h}_{50}^{-1}{\rm Mpc})\approx 5\times 10^{14}h_{50}^{-1}~M_{\sun}$. If the mass
of RXJ0848.9+4452 is not less than the latter (based on its higher
$L_X$), one would conclude that the mass of the supercluster is
$\sim\! 10^{15}h_{50}^{-1}~M_{\sun}$. This estimate is very uncertain
however, because we do not know the virialization state of the two
systems.

The existence of a superstructure at such large look-back times, when
compared with recent results of strong clustering of galaxies at much
higher redshift (Steidel et al.\ 1997), can provide some clues on the
epoch of cluster formation. Strong concentrations of Lyman break
galaxies (LBG) are now found to be ubiquitous in the field at $z\sim\!
3$ (Steidel et al.\ 1998a), with one prominent peak in the redshift
distribution every $\sim\! 100$ arcmin$^2$ (Steidel et al.\ 1998a;
Giavalisco, private communication). Galaxies belonging to these peaks are not
spatially concentrated on the sky and hence are believed to be
non-virialized structures prior to collapse.  Using the effective
survey volume of the LBG survey (Steidel et al.\ 1998b), we can then
estimate the comoving space density of these concentrations at $z\approx 3$ to
be $5\times 10^{-7} {\rm h}_{50}^{3}{\rm Mpc}^{-3}$ and $2\times
10^{-6} {\rm h}_{50}^{3}{\rm Mpc}^{-3}$ for $\Omega=0.2$ and 1
respectively ($\Lambda=0$).  The comoving volume explored by our X-ray cluster
search in the Lynx field (0.2 deg$^2$) for clusters of $L_X\ge
10^{44}\lum$ is $(1-2)\times 10^6 {\rm Mpc}^3{\rm h}_{50}^{-3}$ and
thus, it is reasonable to argue that the cluster pair we have
identified at $z=1.27$ is the evolved version of two (or more) LBG
concentrations at $z\simeq 3\pm 0.4$, well after the collapse, and
possibly in the process of merging. Several clusters identified at
$z\simeq 0.8$ in the RDCS or other X-ray surveys (e.g.,\ Gioia et al.\
1998) show a filamentary structure, often with two cores (both in
optical and X-ray) separated by approximately 1 Mpc comoving.  By
comparing the space density of LBG concentrations with the local
cluster abundance, Steidel et al. (1997), and Governato et al. (1998)
on more quantitative grounds, have argued that these high redshift
concentrations are likely the progenitors of local rich clusters. In
this respect, we note that the space density of local X-ray clusters
more luminous than the Lynx clusters is $N(L_X>10^{44}\lum) \simeq
3\times 10^{-7} {\rm h}_{50}^{3}{\rm Mpc}^{-3} $ (e.g., Ebeling et al. 1997).

\subsection{The cluster galaxy population}

The color-magnitude diagrams in Figure~\ref{f:cmdiagram} show the
presence of a red envelope which in lower $z$ clusters is  dominated by early-type galaxies (SED98).  
Also plotted in the panels are estimates of the
no--evolution color--magnitude locus for early--type galaxies.  The
dashed lines were calculated as described in SED98 using photometry of
early--type galaxies from a $UBVRIzJHK$ dataset covering the central
$\sim$1 Mpc of the Coma cluster (Eisenhardt et al.\ 1999).  These data
enable us to determine, by interpolation, the colors that Coma
galaxies would appear to have if the cluster could be placed at $z =
1.26$ and observed through the $RIJK$ filters used on RXJ0848.9+4452.
The same physical apertures were used to measure colors in Coma and in
the two distant clusters.

The colors of the spectroscopic members are broadly similar to those
member galaxies in ClG J0848+4453, with 
the exception of ID\#3 which is considerably bluer in the 
observed colors.  As discussed in S97, the optical-IR colors
are consistent with a passively evolving elliptical model
constructed from a 0.1 Gyr burst population with solar metallicity and
a Salpeter IMF formed at $z = 5$ for $h=0.65$, $\Omega_0 = 0.3$.  The
colors of the red sequences as well as the spectral energy
distributions of the galaxies in the two clusters are also similar, as
would be expected if the constituent stellar populations formed at
such high redshifts with little subsequent activity either due to
merging or starbursts.  

To more fully address the evolution of galaxy popoulations at these
high redshifts, a larger sample of member galaxies which is complete to 
a given K-band magnitude is needed, along with morphological information.
This would extend the sample to include bluer, star-forming galaxies which are found to be numerous in clusters at $z \approx 0.9$ (Postman et al. 1998),
and which are within reach of optical and near-IR spectrometers on 8-10m
class telescopes.

\section{Conclusions}

Near-infrared imaging of a faint extended X-ray source
(RXJ0848.9+4452) detected in a deep ROSAT-PSPC observation of the Lynx
field has revealed a significant overdensity of red objects with $R-K$
and $J-K$ colors typical of ellipticals at $z>1$. Spectroscopic
observations with Keck-LRIS have secured redshifts for 6 galaxies in
the range $z=1.262\pm 0.005$ within a 35\arcsec\ radius region around
the peak of the X-ray emission. These data indicate the presence
of a moderately rich galaxy cluster at $<z>=1.261$ with rest frame
X-ray luminosity $L_X \simeq 1.5\times 10^{44}$ ergs s$^{-1}$
(0.5--2.0 keV band).  RXJ0848.9+4452 is the highest redshift X-ray
selected cluster found to date.

An interesting circumstance is that this system lies only 4.2 arcmin
away from ClG~J0848+4453, an IR-selected cluster previously discovered
by Stanford et al. (1997) at $<z>=1.273$, also known to be X-ray
luminous with half the $L_X$ of RXJ0848.9+4452. Ten spectroscopic
members are known to date in ClG~J0848+4453 and several others have
been found in a 28 arcmin$^2$ field surrounding these two systems. We
therefore find evidence of a high redshift superstructure,
consisting of two separate systems, in an advanced stage of collapse
as elucidated by the X-ray data. The two cluster cores are separated by
$5\, {\rm h}^{-1}_{50}$ comoving Mpc. Furthermore, the redshift
distribution of all the spectroscopic members suggests a velocity
difference between the two clusters $c\Delta z/(1+z) \lesssim 1500$
km/s, leading us to speculate that these systems are in the process
of merging.

The member galaxies in the two clusters form a homogeneous population
with similar colors, the difference in the median value of $J-K$ being
less than 0.05 mag. This supports the conclusion of S97 and SED98 who
found that the spectrophotometric properties of these red galaxies are
consistent with passively-evolving ellipticals formed at high
redshift.

This work shows that the combination of near-IR imaging and the X-ray
selection from deep X-ray pointings can lead to the identification of
bona-fide clusters at $z>1$. The additional advantage of this approach
is that the X-ray selection allows the search volume to be estimated
and thus the cluster space density to be effectively evaluated, along
with the cluster mass function once the cluster mass is measured.  The
current limitation of this approach is due to the limited resolution
and sensitivity of current X-ray observations which require pushing
these studies to flux levels where confusion becomes important and the
effective sky coverage becomes very small.  With the advent of AXAF and
XMM, the use of this strategy for cluster searches at $z>1$ promises to
be extremely effective, possibly with a success rate of identification
as high as the one obtained with the ROSAT-PSPC at $z<1$. A census of
galaxy clusters at $1<z<1.5$ will be a major breakthrough toward our
understanding of the formation of both clusters and their constituent
galaxy populations.

\acknowledgments

PR aknowledges partial support from NASA ADP grant NAG 5-3537 and
thanks Roberto Della Ceca, Riccardo Giacconi and Colin Norman for their
continuous support on the RDCS work.  
We thank Tom Jarrett for supporting observations with the Palomar Prime
Focus Infrared Camera, which was a pleasure to use.
Portions of this work were carried out by the Jet Propulsion Laboratory,
California Institute of Technology, under a contract with NASA.
Part of the observational material
presented here was obtained at the W.\ M.\ Keck Observatory, which is a
scientific partnership between the University of California and the
California Institute of Technology, made possible by a generous gift of
the W.\ M.\ Keck Foundation.  The work by SAS at LLNL was performed
under the auspices of the U.S. Department of Energy under Contract No.\
W-7405-ENG-48. AD acknowledges the support of NASA HF-01089.01-97A.

\newpage

\newpage

\newpage

\newpage

\newpage

\newpage

\newpage

\begin{figure}
\epsscale{1.05}
 \vspace*{15cm}
\caption{A composite $BIK$ image of Lynx field covering the same area as in Figure~\protect\ref{f:xcont_obj}. The inset is a blowup of the core 
of RXJ0848.9+4452.}
\label{f:bik}
\end{figure}
\newpage

\newpage

\clearpage 

\begin{deluxetable}{lcccccccc}
\small
\tablenum{1}
\tablecolumns{9}
\tablewidth{0pc}
\tablecaption{X-ray Cluster Parameters in Lynx field}
\tablehead{
  \colhead{ROSAT ID}
& \colhead{RA\tablenotemark{a}}
& \colhead{DEC}
& \colhead{$\theta$\ \tablenotemark{b}}
& \colhead{Counts\tablenotemark{c}}
& \colhead{$F_X$\tablenotemark{d}}
& \colhead{$\sigma_{\rm ext}$\tablenotemark{e}}
& \colhead{Redshift} 
& \colhead{$L_X$\tablenotemark{f}} 
}
\startdata
RXJ0848.8+4456  & 08:48:46.9 & +44:56:22 & 7.4 & $191\pm 18$ & $3.4\pm 0.3$ 
	& 3.5 & 0.571 & $0.50\pm 0.05$ \nl
RXJ0848.9+4452  & 08:48:56.2 & +44:52:00 & 3.2 & $110\pm 17$ & $1.8\pm 0.3$
	& 5.0 & 1.261 & $1.5\pm 0.3$ \nl
RXJ0848.6+4453\tablenotemark{g}
              & 08:48:34.2 & +44:53:35 & 7.5 & $60\pm 16$  & $1.0\pm 0.3$
	& 3.0 & 1.273 & $0.8\pm 0.3$ \nl
\enddata
\tablenotetext{a}{J2000 coordinates}
\tablenotetext{b}{Off-axis angle in arcmin}
\tablenotetext{c}{Source net counts in the [0.5-2.0] keV band}
\tablenotetext{d}{Unabsorbed X-ray flux in units of $(10^{-14}\fl)$, in the [0.5-2.0] keV band in an aperture of 2\arcmin\ radius.}
\tablenotetext{e}{Statistical significance of the source extent
provided by wavelet algorithm}
\tablenotetext{f}{X-ray luminosity in units of $(10^{44}\lum)$, in the 
rest frame [0.5-2.0] keV band ($H_0=50, \, q_0=0.5$ assumed). }
\tablenotetext{g}{alias ClG~J0848+4453, Stanford et al. 1997}
\label{t:xray}
\end{deluxetable}
\normalsize

\begin{deluxetable}{ccccccccc}
\tablenum{2}
\tablecaption{Spectrophotometric properties of targets around RXJ0848.9+4452.}
\tablehead{
\colhead{ID} & \colhead{R.A.} & \colhead{Dec.} & \colhead{$z$} & 
\colhead{$\delta z$} &
\colhead{$K$}  &\colhead{$J-K$} & \colhead{$I-K$} & \colhead{$R-K$}  
}
\startdata

1 &  08:48:58.66 & +44:51:57.2 & 1.257 & 0.001 & 16.72 & 1.86 & 4.16 & 5.91 \nl
11 & 08:48:58.33 & +44:51:55.8 & 1.266 & 0.004 & 17.94 & 1.83 & 4.21 & 5.90 \nl
3  & 08:48:59.07 & +44:51:44.3 & 1.261 & 0.003 & 18.38  & 1.60 & 3.64 & 5.10 \nl
18 & 08:48:58.57 & +44:51:33.3 & 1.257 & 0.003 & 18.3~  & 1.85 & 4.38 & 6.03 \nl
4  & 08:48:58.87 & +44:51:26.9 & 1.267 & 0.003 & 17.61  & 1.92 & 4.08 & 5.26 \nl
248 & 08:48:55.94 & +44:51:55.1 & 1.254 & 0.005 & 19.25 & 1.67 & 4.37 & 5.60\nl
\hline
20 & 08:49:02.11 & +44:51:09.0 & 0.818 & 0.003 & 18.78  & 1.63 & 3.50 & 4.68 \nl
8  & 08:48:56.03 & +44:51:30.9 & 1.144 & 0.003 & 17.72 & 1.76 & 3.78 & 5.55 \nl
2  & 08:48:57.28 & +44:51:58.2 & \nodata & \nodata & 19.09 & 1.60 & 3.02 & 4.06 \nl
228 & 08:48:56.68 & +44:51:55.8 & \nodata & \nodata & 19.15 & 1.99 & 4.62 & 5.76 
\enddata
\end{deluxetable}

\end{document}